\documentstyle[11pt,epsfig,here,a4]{article}
\begin{document}

\vspace*{3mm}
\begin{center}
{\Large \bf The Baikal Deep Underwater Neutrino Experiment:}

{\Large \bf Results, Status, Future $^*$} 
\end{center}

\begin{center}
{\sc Ch.Spiering} \\
{\small for the  BAIKAL Collaboration:}
\end{center}
\vspace{-2mm}
{\footnotesize
V.A.Balkanov$^2$, I.A.Belolaptikov$^7$, L.B.Bezrukov$^1$, N.M.Budnev$^2$, 
A.G.Chensky$^2$, I.A.Danilchenko$^1$, Zh.-A.Djilkibaev$^1$, 
G.V.Domogatsky$^1$, A.A.Doroshenko$^1$, S.V.Fialkovsky$^4$, O.N.Gaponenko$^2$, 
A.A.Garus$^1$, T.I.Gress$^2$, A.M.Klabukov$^1$, A.I.Klimov$^6$, 
S.I.Klimushin$^1$, A.P.Koshechkin$^1$, V.F.Kulepov$^4$, L.A.Kuzmichev$^3$, 
S.V.Lovzov$^2$, B.K.Lubsandorzhiev$^1$, M.B.Milenin$^4$, R.R.Mirgazov$^2$, 
A.V.Moroz$^2$, N.I.Moseiko$^3$, S.A.Nikiforov$^2$, E.A.Osipova$^3$, 
A.I.Panfilov$^1$, Yu.V.Parfenov$^2$, A.A.Pavlov$^2$, D.P.Petukhov$^1$, 
P.G.Pokhil$^1$, P.A.Pokolev$^2$, E.G.Popova$^3$, M.I.Rozanov$^5$, 
V.Yu.Rubzov$^2$, I.A.Sokalski$^1$, Ch.Spiering$^8$, O.Streicher$^8$, 
B.A.Tarashansky$^2$, T.Thon$^8$, R.Wischnewski$^8$, I.V.Yashin$^3$

{\it 1 - Institute  for  Nuclear  Research,  Russian  Academy of Science
(Moscow); \mbox{2 - Irkutsk} State University (Irkutsk); \mbox{3 - Moscow}
State University (Moscow); \mbox{4 - Nizhni}  Novgorod  State  Technical
University  (Nizhni   Novgorod);\\ 5 - St.Petersburg State Marine
Technical  University (St. Petersburg); \mbox{6 - Kurchatov} Institute
(Moscow);\\ \mbox{7 - Joint} Institute for Nuclear Research (Dubna);
\mbox8 - DESY-Zeuthen (Zeuthen) }}

\vspace{0.5cm}

{\footnotesize
We review the present status of the Baikal Underwater Neutrino 
Experiment and present results obtained
with the various stages of the stepwise increasing detector: {\it NT-36} 
(1993-95), {\it NT-72} (1995-96) and {\it NT-96} (1996-97).
Results cover atmospheric muons, 
first clear neutrino events, search for neutrinos from WIMP
annihilation in the center of the Earth, search for magnetic
monopoles, and -- far from astroparticle physics -- limnology.}

\section{Detector and Site}

The Baikal Neutrino Telescope is being deployed in Lake Baikal, Siberia, 
\mbox{3.6 km} from shore at a depth of \mbox{1.1 km} (see fig.1). 
At this depth, the light absorbtion length for wavelengths between
470 and \mbox{500 nm} is about 20 m.
Scattering is strongly forward peaked ($\langle \cos \theta
\rangle) \approx 0.95)$. Typical values for the scattering length 
are about 15 m.
Expressed in terms of the {\it effective} scattering length
$L_{eff} = L_{scatt} / (1 - \langle \cos \theta
\rangle)$, this corresponds to $L_{eff}$ = 300\,m.

{\it NT-200}, the medium-term goal of the collaboration
\cite{APP, Proj}, will
be finished in April 1998 and will consist of 192
optical modules (OMs). An umbrella-like frame carries  8 strings,
each with 24 pairwise arranged OMs.
Three underwater electrical cables connect the
detector with the shore station. Deployment of all detector 
components is carried out  during 5--7 weeks in late winter when 
the lake is covered by thick ice.

In April 1993, the first part of {\it NT-200}, the detector {\it
NT-36} with 36 OMs at 3 short strings, was put into operation 
and took data up to March 1995. A 72-OM array, {\it \mbox{NT-72}}, 
run in \mbox{1995-96}. In 1996 it
was replaced by the four-string array {\it NT-96}. Summed over 700
days effective life time, $3.2\cdot 10^{8}$ muon events have been 
collected with
\mbox{{\it NT-36, -72, -96}}. Since \mbox{April 6,} 1997, {\it
  NT-144}, a six-string array with 144 OMs, is taking data (see fig.2).

The OMs are grouped in pairs along the strings. They contain 
37-cm diameter {\it QUASAR} PMTs. The two PMTs of a
pair are switched in coincidence in order to suppress background
from bioluminescence and PMT noise.
A pair defines a {\it channel}.

\vspace{0.2cm}

\begin{center}
{\large \it * Talk given at the Int. School of 
Nuclear Physics, Erice, Sept.1997}
\end{center}

\begin{figure}[H]
\centering
  \mbox{\epsfig{file=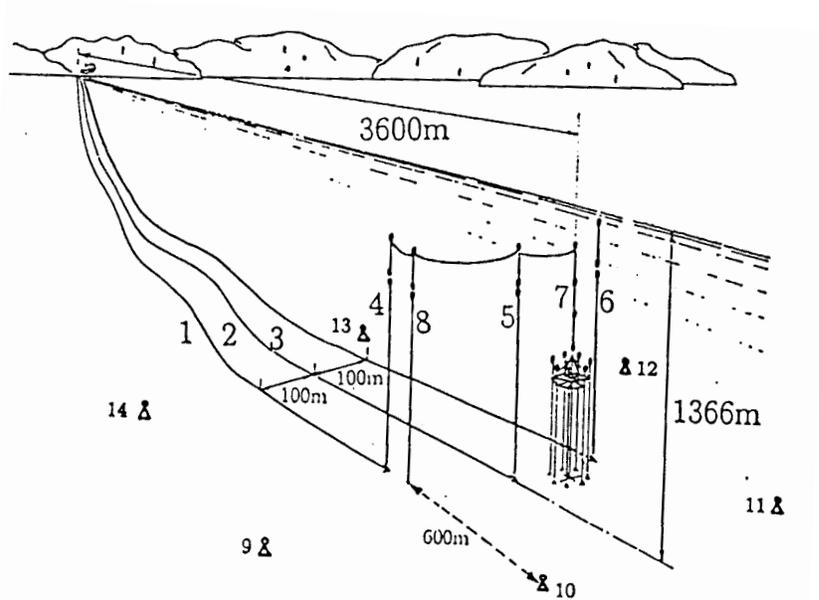,height=11cm, angle=87.5}}
\caption[2]{\small
 The Baikal detector complex (status since 1994).
                  {\it 1,2} -- wire cables to shore,
                  {\it 3} -- opto-electrical cable to shore,
                  {\it 4,5,6} -- string stations for shore cables
                  1,2,3,
                  respectively,
                  {\it 7} -- string with the telescope,
                  {\it 8} -- hydrometric string,
                  {\it 9-14}  -- ultrasonic emitters
}
\end{figure}

\vspace{-1.7cm}

\begin{figure}[H]
\centering
  \mbox{\epsfig{file=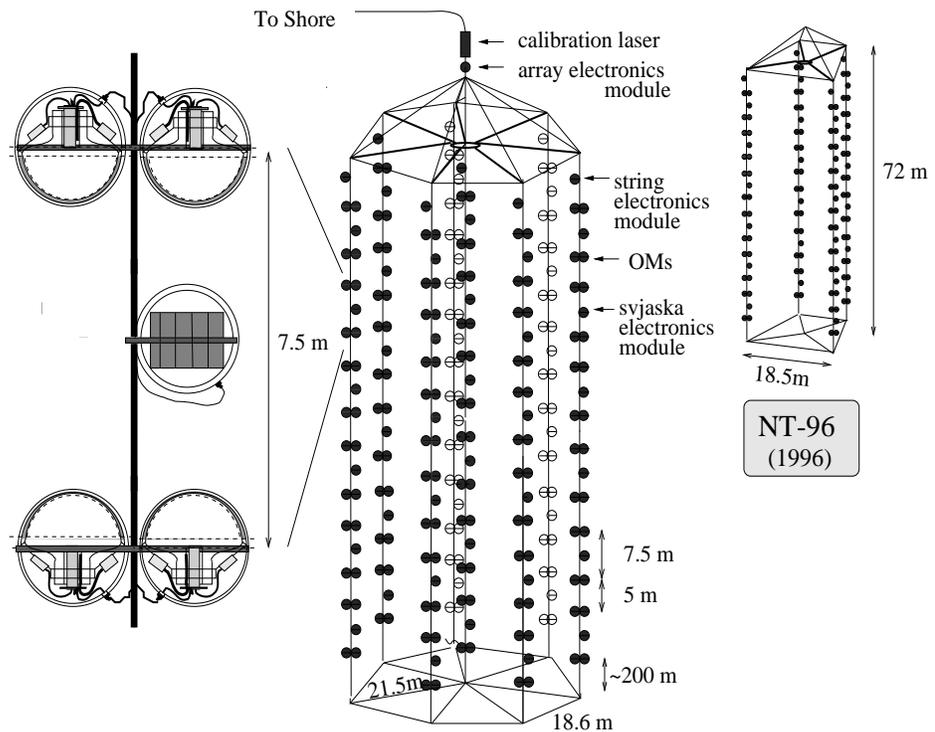,height=13cm}}
\vspace{-1cm}
\caption[4]{\small
Schematic view of the Baikal Telescope {\it NT-200}.
                  The modules of {\it NT-144}, operating since April
                  1997, are in black.
                  The expansion left-hand shows 2 pairs of
                  optical modules ("svjaska") with the svjaska
                  electronics module, which houses
                  parts of the read-out and control electronics.
                   Top right the array {\it NT-96} is shown,
                  which took data between April 1996 and March 1997.
}
\end{figure}

 A {\it muon-trigger}
is formed by the requirement of \mbox{$\geq N$ {\it hits}}
(with {\it hit} referring to a channel) within \mbox{500 ns}.
$N$ is typically set to the value
\mbox{3 or 4.} For  such  events, amplitude and time of all fired
channels are digitized and sent to shore. The event record
includes all hits within a time
window of -1.0 $\mu$sec to +0.8 $\mu$sec with respect to the muon
trigger signal.
A separate {\em monopole trigger} system searches for clusters of
sequential hits in individual channels which are 
characteristic for the passage of slowly moving, bright
objects like GUT monopoles. 

In the initial project of {\it NT-200}, the optical modules were 
directed alternately upward and downward, with a 
distance of 7.5\,m between pairs looking face to face, and of 5\,m
between pairs arranged 
back to back.
Due to sedimentation of biomatter deterioiating the
sensitivity of upward looking OMs we were forced to
direct  the OMs of the present arrays essentially downward.

\section{Track Reconstruction}

The parameters of a muon 
track crossing the detector are determined by 
minimizing \cite{Reco}

\vspace{-2mm}
\begin{equation}
\chi^2_t = \sum_{i=1}^{N_{hit}} (T_i(\theta, \phi, u_0, v_0, t_0)
    - t_i)^2 / \sigma_{ti}^2
\end{equation}
\vspace{-2mm}

\noindent
Here, $t_i$ are the measured times and $T_i$ the times expected for a given 
set of track parameters and simplifying the theoretical
picture to that of a "naked" muon not
accompanied by electromagnetic showers. 
$N_{hit}$ is the number of hit channels, 
$\sigma_{ti}$ are the timing errors. A set of parameters defining a straight 
track is given by $\theta$ and $\phi$ -- zenith and azimuth angles of the 
track, respectively, $u_0$ and $v_0$ -- the two coordinates of the track point 
closest to the center of the detector, and $t_0$ -- the time the muon passes 
this point. 

Only events fulfilling the condition  \mbox{``$\geq 6$} hits at
\mbox{$\geq 3$} strings`` (trigger {\it 6/3})
are selected for the standard track reconstruction 
procedure which consists of the following steps:
\vspace{-2mm}
\begin{enumerate}
\item A preliminary analysis including several causality criteria 
which reject 
events  violating the model of a naked muon. After that, a 0-th 
approximation of $\theta$ and $\phi$ is performed.
\vspace{-1mm}
\item A minimum search (minimization of $\chi^2_t)$, 
based on the model of a naked muon.
\vspace{-1mm}
\item Quality criteria to reject most badly reconstructed events.
\end{enumerate}
\vspace{-2mm}
The reconstruction procedure is described in detail in \cite{APP,Reco}. 
Fig.4
shows a typical single muon  event
firing 7 of the 18 channels of {\it NT-36} and
reconstructed with a $\chi^2/NDF = 0.57$ ({\it left}),
as well as  the muon intensity $I_{\mu}(\theta)$ as a function of 
measured zenith angle  at the detector depth ({\it right}). 
The angular distribution is well described by MC expectations.

\section{Separation of Neutrino Events}

The canonical signature of neutrino induced events is a muon crossing
the detector from below.
With the flux of downward muons exceeding that of
upward muons from atmospheric neutrino interactions by
 about 6 orders of magnitude, a careful reconstruction is of
prime importance.
Clear neutrino signals in the rather small {\it NT-36} could be separated
only over a
very limited cone around the opposite zenith. Contrary to that,
{\it NT-96} can be
considered as a real  neutrino telescope for a wide region in
zenith angle $\theta$.

\begin{figure}[H]
\centering
  \mbox{\epsfig{file=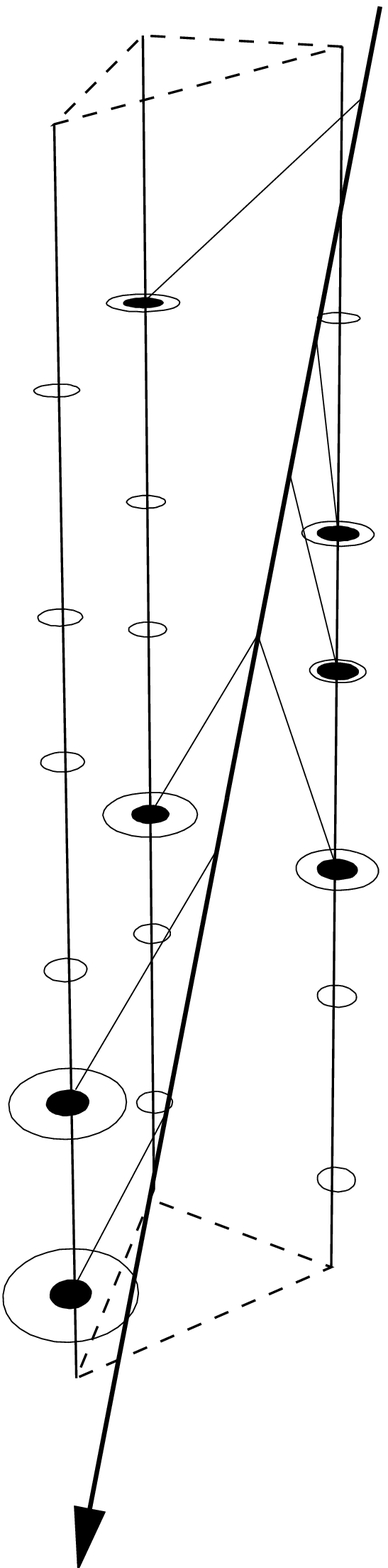,height=7cm,width=3.5cm}}
\hspace{2.5cm}
  \mbox{\epsfig{file=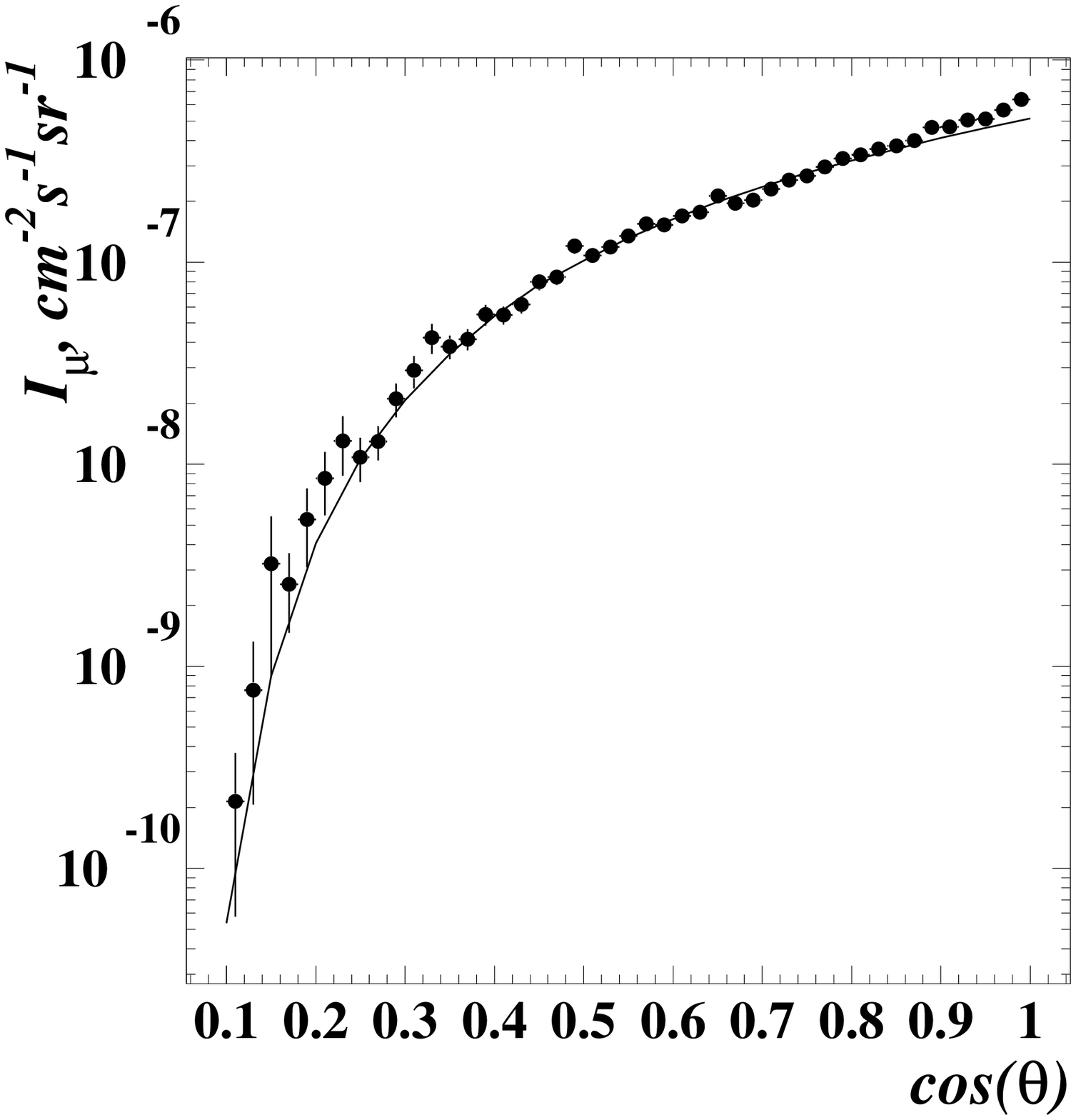,height=7cm,width=7cm}}
\caption[15]{\small
{\it Left:} Single muon event recorded with {\it NT-36}.
Hit channels are in black. The thick line gives the
reconstructed path, thin lines pointing to the channels mark
the path of Cherenkov photons as given by the fit to the
measured times.
The sizes of the ellipses are proportional to the
recorded amplitudes.
{\it Right:}
Zenith angle distribution of the muon
         intensity $I_{\mu}(\theta)$ at a depth of 1170 m. Full
         circles:
         experimental data, solid line: $I_{\mu}(\theta)$
         calculated for stochastic energy loss.
}
\end{figure}

\subsection{Search for Nearly Upward Moving Neutrinos with NT-36}
 
For separation of nearly vertical upward muons in the {\it NT-36} prototype
array we used special
separation criteria instead of full reconstruction (see also
\cite{ourneu}). These criteria
make use of two facts: firstly, that the muons searched for have
the same  vertical direction like the string; secondly, that
low-energy muons generate mainly direct Cherenkov light and, 
consequently, are not visible over
large distances and should produce a clear time and amplitude pattern
in the detector -- mostly only along {\it one} string. 
We have chosen the following criteria:

\begin{enumerate}

\item  {\large $ |(t_{i}-t_{j})-(T_{i}-T_{j})|<dt$}:
{\bf the time pattern must be close to that of a straight
upward moving muon.}
$t_i(t_j)$ are the measured times in any hit 
channels $i(j)$, $T_i(T_j)$ are
the ``theoretical'' times expected for naked, up-going
vertical muons and $dt$ is a time cut.

\item {\large $ dA_{ij}(down-up) > 0.3 $}:
{\bf All down-looking channels must see clearly more light
than any upward looking channel.}
$dA_{ij}(down-up)=(A_{i}(down)-A_{j}(up))/(A_{i}(down)+A_{j}(up))$ and
$A_{i}(down) (A_{j}(up))$ are the amplitudes of channel $i(j)$ facing
downward (upward).

\item {\large $ A_i(down)$$>$$4 pe $}: 
{\bf Amplitudes of downward channels must exceed 4
photoelectrons}

\item {\large $ dA(down-down)<0. $}
{\bf The light must not decrease from bottom to top, as one
expects for showers generated by downward muons below the array.}
$dA(down-down)$ is defined as that of the 3 possible combinations
$dA_{ij}(down-down)=(A_i-A_j)/(A_i+A_j)\mid _{i>j}$ of
downward channels which has the largest
absolute value. For events due to showers below the array 
it peaks at values close to 1, for vertical neutrino candidates it 
should be close to zero. The criterium rejects nearly all shower
events but only half of the neutrino sample.

\end{enumerate}

We analyzed the data taken with {\it
  \mbox{NT-36}}
between April 8, 1994 and March 5, 1995 (212 days lifetime). There
were 6 PMT
pairs along each of the 3 strings. The orientation of the
channels
from top to bottom  at each string was
{\em up-down-up-down-up-down}. Upward-going muon candidates
were selected from a total of \mbox{$8.33\cdot 10^{7}$} events 
recorded by the muon-trigger \mbox{"$\geq 3$} hit channels". 
The samples fulfilling trigger
conditions {\it 1, 1-2, 1-3} and {\it 1--4} with time cut $dt$\,=\,20\,ns
contain  131, 17 and 2 events, respectively. 
Only two events fulfill trigger conditions
{\it 1--3} and {\it 1--4}. 
The first is consistent with a nearly vertical 
upward going muon
and the second  with an upward going muon with zenith angle
$\theta_{\mu}=15^{\circ}$ (fig.4). A detailed analysis \cite{ourneu}
yields a fake probability of $\le$3\% for both events.

\begin{figure}[H]
\centering
  \mbox{\epsfig{file=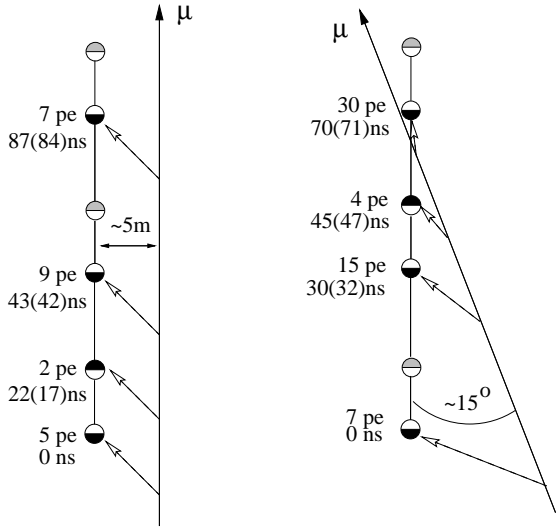,height=7.3cm,angle=-90}}
\caption[12]{\small
The two neutrino candidates in NT-36. The hit PMT pairs
(channels) are
marked in black. Numbers give the measured amplitudes (in
photoelectrons) and
times with respect to the first hit channel. Times in brackets are
those expected for a vertical going upward muon ({\it left}) 
and an upward muon passing
the string under \mbox{$15^o$} ({\it right}).
}
\end{figure}

Considering the two neutrino candidates as atmospheric neutrino
events, a 90 \% CL upper limit of $1.3 \cdot 10^{-13}$ 
(muons cm$^{-2}$ sec$^{-1}$) in
a cone with 15 degree half-aperture around the opposite zenith is
obtained for upward going muons generated by neutrinos 
due to neutralino annihilation in the
center of the Earth. The limit corresponds to muons with energies
greater than  the threshold energy $E_{th} \approx 6$ GeV, 
defined by 30m string length.
This is still an order of magnitude higher than the limits 
obtained by Kamiokande~\cite{kamneutralino}, 
Baksan~\cite{bakneutralino} and MACRO~\cite{MACROneutralino}. 
The effective area of {\it NT-36} for nearly
vertical upward going muons fulfilling our separation criteria 
{\it 1-3} with $dt$=20\,ns is $S_{eff}=50$ m$^{2}$/string. 
A rough estimate of the effective
area of the full-scale {\it \mbox{NT-200}}
(with  eight strings twice as long as those of 
{\it \mbox{NT-36}}) with respect to
nearly vertical upward muons gives $S_{eff} 
\approx 400-800$ m$^{2}$.

\subsection{Fully reconstructed neutrinos separated with NT-96}

The {\it NT-96} data are analyzed using the standard reconstruction 
procedure described in sect.2 as well as a method similar to that
described in the previous subsection. In this subsection we
present results obtained with the standard procedure 
(see also \cite{neutrino97}).
For {\it NT-96}, the most effective quality cuts are the traditional 
$\chi^2$ cut, cuts on the probability of non-fired channels not to be hit, and 
fired channels to be hit ($P_{nohit}$ and $P_{hit}$, respectively), cuts on 
the correlation function of measured amplitudes to the amplitudes expected for
the reconstructed tracks, and a cut on the amplitude $\chi^2$ defined similar 
to the time $\chi^2$ defined above. To guarantee a minimum lever arm for track 
fitting, we reject events with a projection of the most distant channels on 
the track ($Z_{dist}$) below 35 meters. Due to the small transversal 
dimensions of {\it NT-96}, this cut excludes zenith angles close to the 
horizon, i.e., the effective area of the detector with respect to atmospheric 
neutrinos is decreased considerably (fig.5, left).

The efficiency of all criteria has been tested using MC generated atmospheric
muons and upward muons due to atmospheric neutrinos. $ 1.8 \cdot 10^6$ events 
from atmospheric muon events (trigger {\it 6/3}) have been simulated, with
only 2 of them passing all cuts and being reconstructed as upward going muons.
This corresponds to $S/N \approx 1$. Rejecting all events with less than 9 
hits, no MC fake event is left, with only a small decrease in neutrino 
sensitivity. This corresponds to $S/N > 1$ and the lowest curve in fig.5.

With this procedure, we have reconstructed $5.3 \cdot 10^6$ events taken with 
{\it NT-96} in April/May 1996. The resulting angular distribution is presented 
in fig.5 (right). 


\begin{figure}[H]
\centering
  \mbox{\epsfig{file=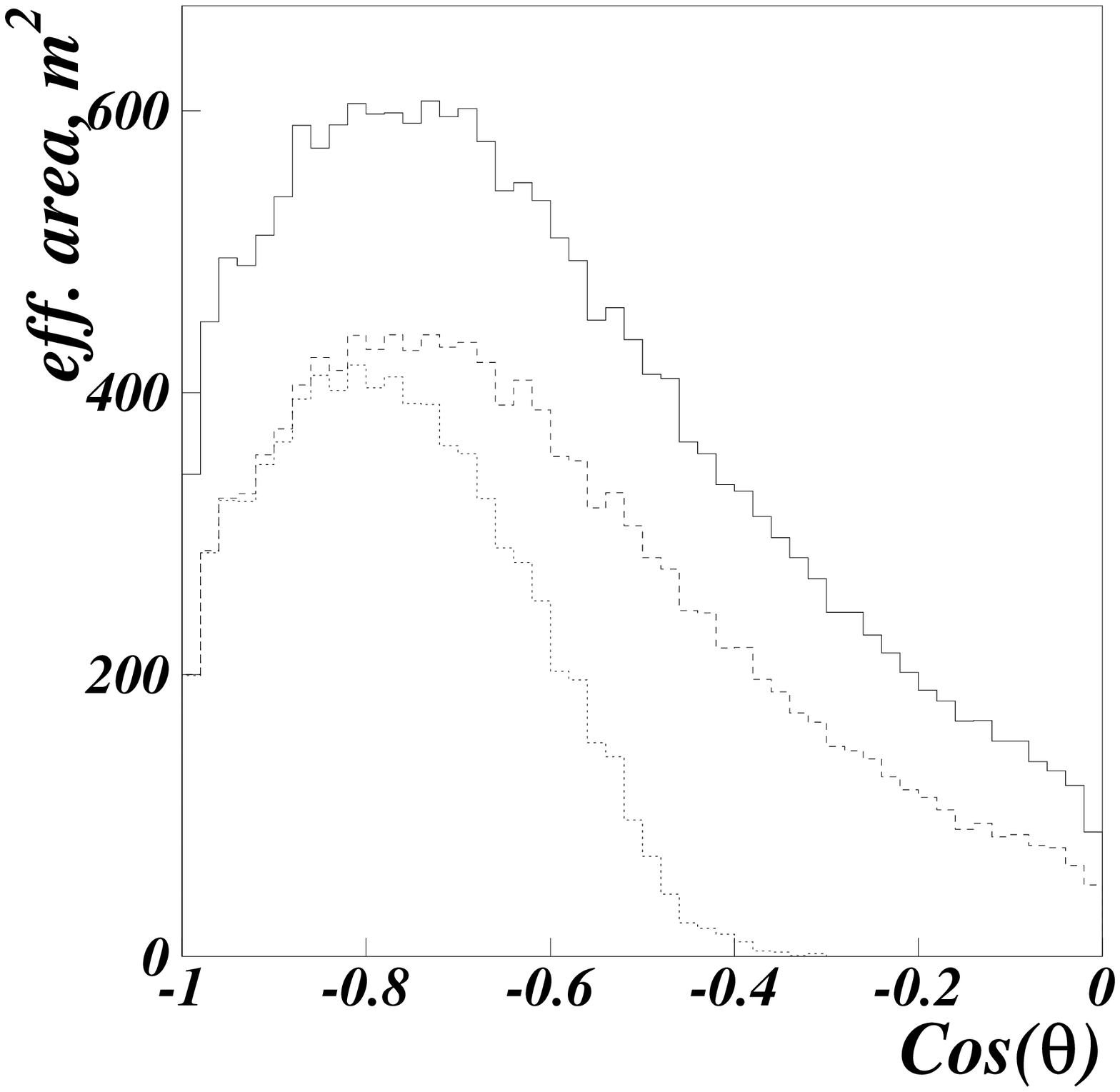,height=7.5cm}}
  \mbox{\epsfig{file=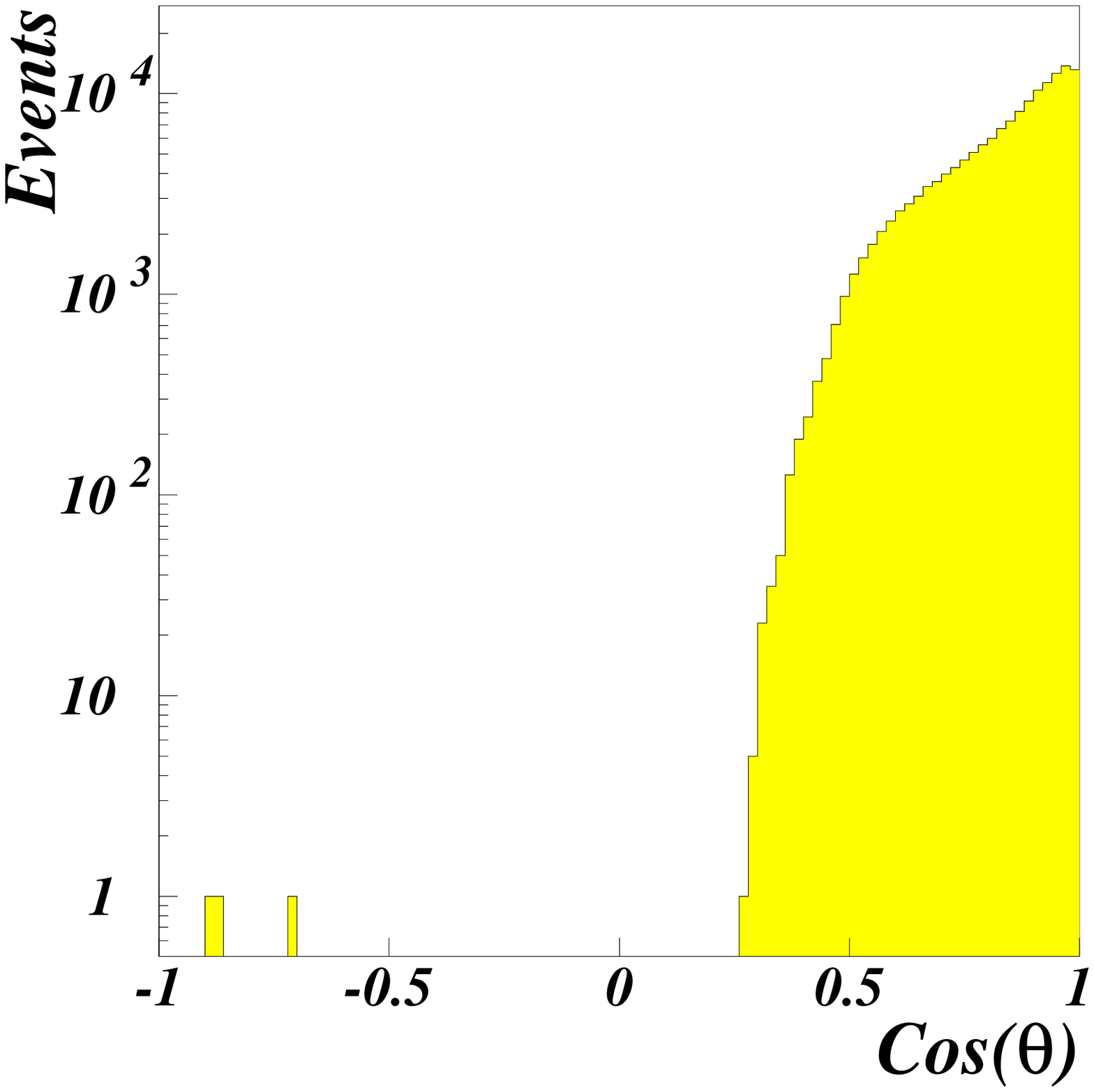,height=7.5cm}}
\caption[12]{\small
{\it Left:}  Effective area for upward muons satisfying trigger {\it 9/3};
solid line -- no quality cuts; dashed line -- final quality cuts;
dotted line
-- final quality cuts and restriction on $Z_{dist}$ (see text).
{\it Right:} Experimental angular distribution of events satisfying trigger
{\it 9/3}, all final quality cuts and the limit on $Z_{dist}$ (see text).
}
\end{figure}

From the time period between April 16 and May 17, 1996 (18 days
lifetime), three neutrino candidates have been separated (see
fig.5, right), 
in good agreement with the expected number of approximately 2.3.
Fig.6 displays one of the neutrino candidates.
Top right the times of the hit channels are shown as a function of the 
vertical position of the channel. At each string we observe the time 
dependence characteristically for upward moving particles.
The 
angle regions ${\psi^{min} -\psi^{max}}$ consistent with the observed time 
differences $\Delta t_{ij}$  between two channels {\it i}, {\it j} are
given by

\begin{equation}
\label{eq:thetalimit}
\cos(\psi^{min}+\eta) < \cos\psi \frac{c \cdot \Delta t_{ij}}
{\vec{r}_j-\vec{r}_i} < \cos({\psi^{max}-\eta})
\end{equation}

\noindent
with ${\vec{r}_i,\vec{r}_j}$ being the coordinates of the channels, 
$\psi$ the muon angle with respect to
$\vec{r}_j-\vec{r}_i$ and $\eta$ the Cherenkov
angle. The bottom right picture of fig.6 shows that the overlap
region of all channel combinations of this event clearly lay
below horizon.

The same holds for 
the other two events, one of which is shown in fig.7a. Fig.7b, in contrast, 
shows an ambiguous event giving, apart from the upward solution, also a 
downward solution. This event is assigned to the downward sample.

\begin{figure}[H]
\centering
  \mbox{\epsfig{file=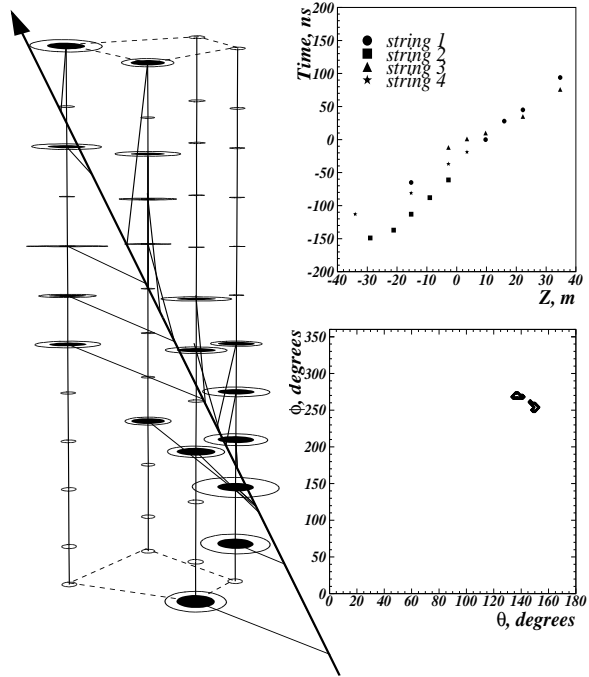,height=11cm}}
\vspace{-1cm}
\caption[12]{\small
 A "gold plated" 19-hit neutrino event. {\it Left:} Event
display. Hit channels are in black. The thick line gives the
reconstructed
muon path, thin lines pointing to the channels mark the path of the
Cherenkov
photons as given by the fit to the measured times. The areas of the
circles
are proportional to the measured amplitudes. {\it Top right:} Hit
times versus
vertical channel positions. {\it Bottom right:}  The allowed
$\theta/\phi$
regions (see text). The fake probability of this event is
smaller than 1\%.
}
\end{figure}

\vspace{-0.6cm}

\begin{figure}[H]
\centering
  \mbox{\epsfig{file=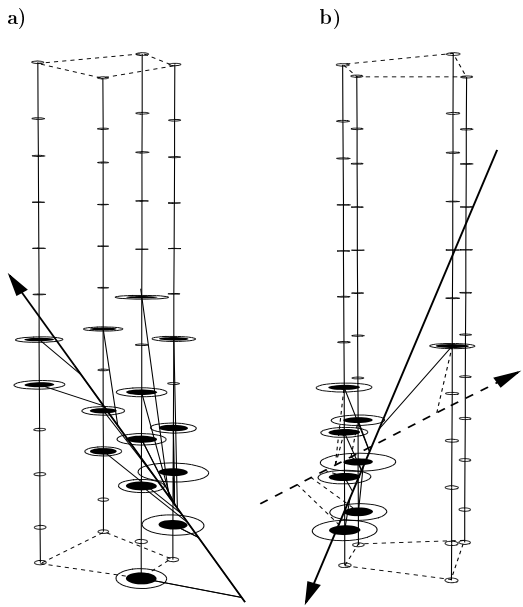,height=9.5cm}}
\vspace{-0.4cm}
\caption[12]{\small
 a) - an unambiguous 14-hit neutrino candidate; b) - an
ambiguous event reconstructed as a neutrino event (dashed line) but
with a
second solution above the horizon (solid line). This event is
assigned to the  sample of downward going muons.
}
\end{figure}

In the mean time, 70 days from {\it NT-96}  have been analyzed, and
12 neutrino candidates have been found. Nine of them have been 
fully reconstructed, 3 nearly upward vertical tracks
hit only 2 strings and give a clear zenith angle but
ambiguities in the azimuth angle -- similar to the two events from {\it
NT-36}. Taking into account the degradation of {\it NT-96} due to failed
OMs, this is in agreement with MC expectations.

{\it NT-200} will have an
effective area of $\approx$\,1500 m$^2$, after all cuts and averaged over
a cone of 60 degrees about the opposite zenith. It will
record about one separable neutrino per day.

\section{Search for Magnetic Monopoles}

GUT monopoles may catalyze processes violating baryon number
conservation~\cite{ruba}. For reasonable velocities $\beta \leq
10^{-3}$, a
cross section \mbox{$\sigma_c = 0.17 \cdot \sigma_o / \beta^{2}$} is
predicted
for monopole-proton interactions~\cite{araf}, with $\sigma_o$ being of
the
order of magnitude typical for strong interactions. The distances
between
sequential proton decays along the monopole track in water can be as
short as
10$^{-2}$ - 10$^{1}$ cm.

In order to search for GUT monopoles, a special trigger was
implemented which
selects events with short-time increase of the counting  rate of
individual
channels, as expected from sequential Cherenkov flashes produced by
the proton
decay products along a monopole track. During standard
data
taking runs, the monopole trigger condition was defined as $\geq3$
hits
within a time window of 500 $\mu$sec in any of the channels.

The data taken from April 16th to November 15th 1993 with {\it NT-36}
were
used to search for monopole candidates.
 We requested that one channel had
counted
\mbox{$\geq7$ hits} and the second channel looking to its face and
located 
\mbox{7.5 m} away along the string \mbox{$\geq3$ hits} during
the same
\mbox{500 $\mu$sec}. This reduces the number of the experimentally
observed
candidates from \mbox{$3.5\cdot 10^{7}$} to zero. From the
non-observation of
monopole candidates we obtain the upper flux limits \mbox{(90 $\%$
  CL)} shown
in Fig.8  together with our earlier results, limits from
IMB~\cite{imbmon} and Kamiokande~\cite{kammon} and with the
astrophysical
Parker limit. Limits of
$10^{-15}$ and
\mbox{$2.7\cdot 10^{-16}$ cm$^{-2}$ s$^{-1}$} 
have been obtained by the MACRO \cite{macmon} and Baksan \cite{bakmon}
experiments, respectively.
With the whole statistics taken since 1993, the
minimal
detectable flux can be lowered by more than an order of
magnitude.

\begin{figure}[H]
\centering
  \mbox{\epsfig{file=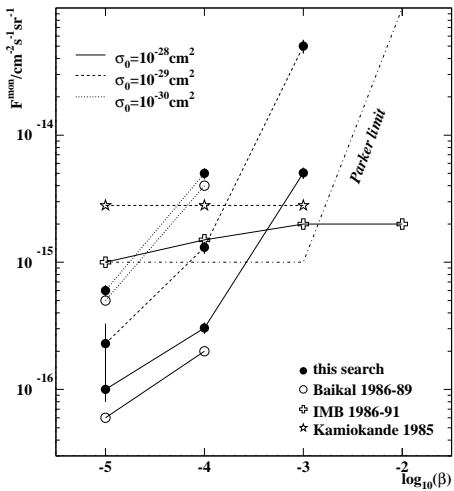,height=6.5cm}}
\caption[2]{\small
Upper limits (90 $\%$ CL) on the flux of magnetic
         GUT monopoles as
         a function of their velocity $\beta$, for different
         catalysis cross sections $\sigma_o$.
}
\end{figure}


Apart from GUT monopoles catalyzing baryon decay, relativistic
monopoles may be detected as well.
The large magnetic charge of a monopole ($g_o = 68.5 e$) 
results in a giant Cherenkov radiation, equal to that  
of  a  14-PeV  muon  for a monopole with $\beta_{mon} \approx 1$. 
The  non-stochastical nature of the Cherenkov emission by relativistic 
monopoles (contrary to a 14 PeV muon!)
may be used to  select monopole candidates. 
Actually,even monopoles with velocities below their Cherenkov
threshold ($\beta_{mon} \approx 0.75$) may be detected, namely by Cherenkov
radiation of $\delta$-electrons ( down to $\beta_{mon} \approx 0.6$).
The effective area of {\it NT-200} for 
monopoles with $\beta \approx 1$ is 
estimated as 20,000 m$^2$.

\section{Limnology}

Apart from its function as a neutrino telescope, the Baikal detector
can be used to monitor  water 
parameters. The array
permanently records photomultiplier counting rates, 
and periodically parameters
like optical transmission at various wavelengths, temperature,
conductivity, pressure (CDT sondes), and speed of sound. 
These measurements form a unique data set 
which can be related to CDT measurements at other locations
in order to build a comprehensive picture of water exchange
processes in the lake. 

For reasons of illustration, 
we show below counting rate variations at various time
scales recorded with  {\it NT-36} in 1993/94.
Counting rates of single PMTs as well as  the "local trigger rates"
(coincidence rates of the 2 PMTs of a pair)
are dominated by  water luminescence.

\begin{figure}[H]
\centering
  \mbox{\epsfig{file=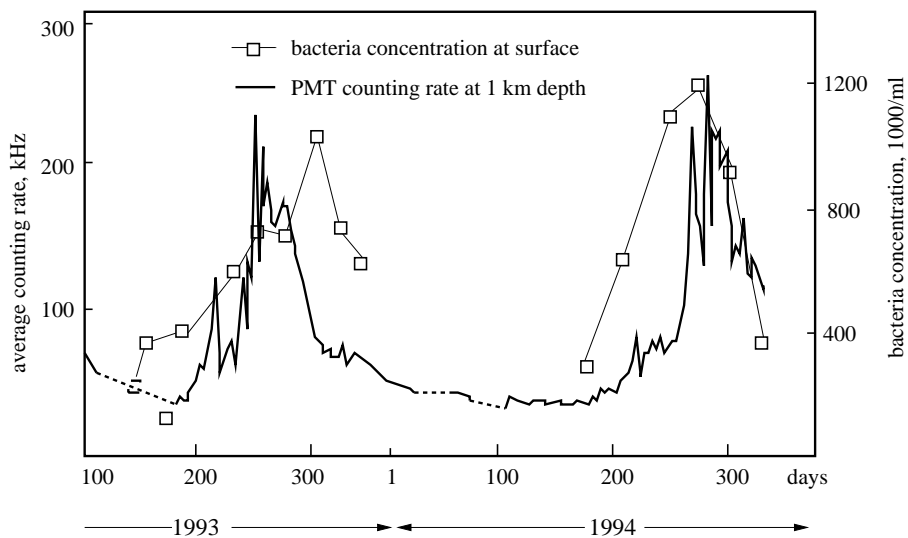,height=12cm,angle=-90}}
\caption[2]{\small
Average counting rate of OMs vs. time, compared to
bacteria concentration at surface.
}
\end{figure}

Fig.9 gives the counting rate over 2 years
and compares it to the bacteria concentration measured at
 a distance of 50 km to the {\it NT-200} site,
at 10 m depth below surface.
In August/September
we observe an increase of the luminosity to extremely high levels.
The changes of the local trigger rate are not
reflected in the muon trigger rate, since the muon trigger is
essentially dominated by atmospheric muons, with negligible
contribution by random hits (water luminescence or dark noise).
This is demonstrated in fig.10 on a shorter time scale,
for a time interval of marked
changes of the local trigger rate following a strong
storm at August 3rd, 1993,
which had washed a lot of water from a nearby river
to the lake.

Fig.11 shows a short period of about 8 hours when the counting
rates sequentially increased, starting with the highest
OMs and ending with the
lowest. From the time shift of the 3 curves a vertical current
of 2.3 cm\,sec$^{-1}$ is deduced. This is noticeable since the
vertical speed of water currents in spring, when the vertical
water renewal is considered to be most intensively, is only
0.2-0.3 cm\,sec$^{-1}$!

\begin{figure}[H]
\centering
\hspace{-6cm}
  \mbox{\epsfig{file=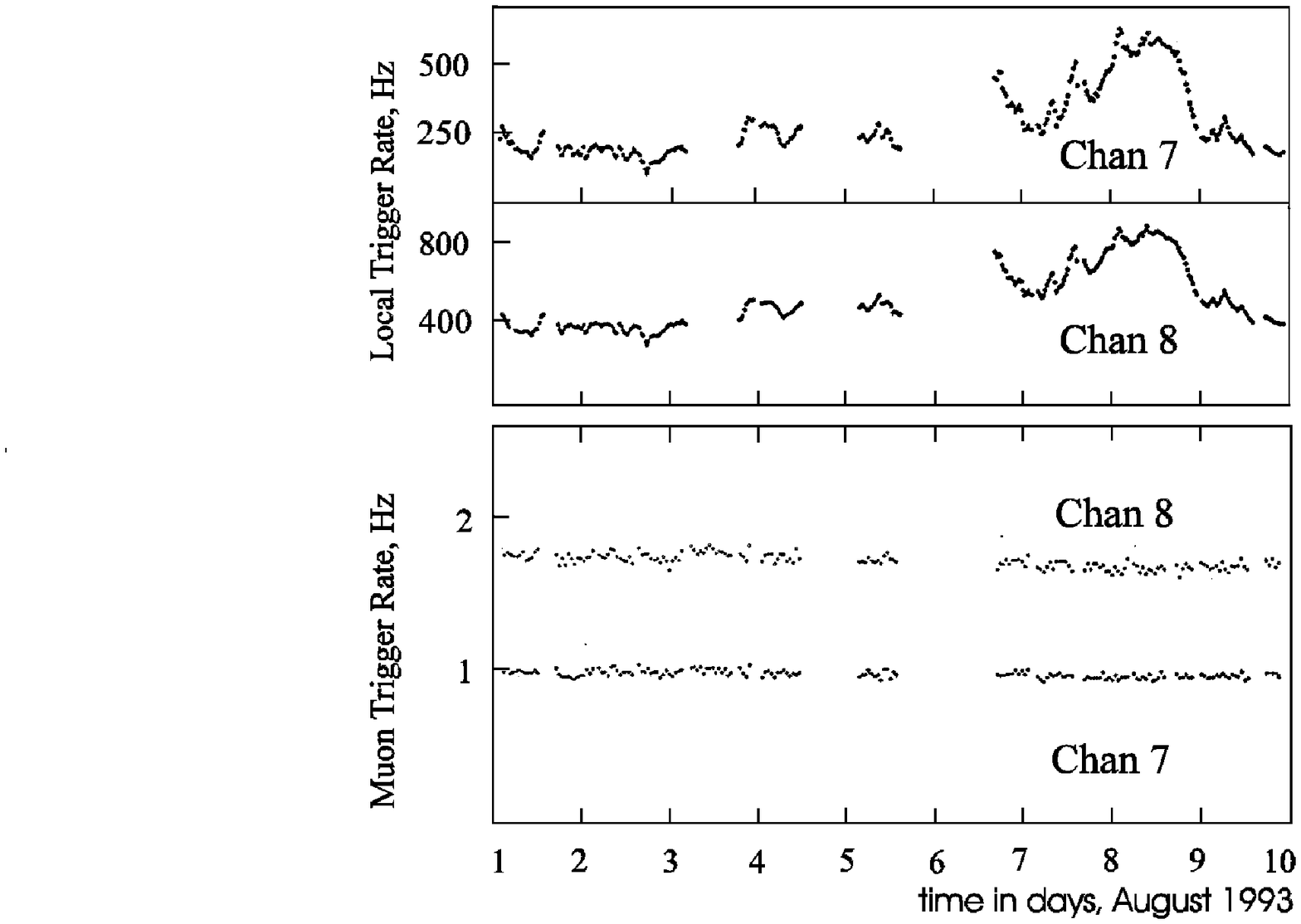,height=8cm,width=12cm}}
\caption[10]{\small
 a) Local trigger rates for channel 7 (downward facing)
            and channel 8 (upward facing) for
            August 1st-9th, 1993. The counting rates are averaged
            over 30\,min.  b) Muon trigger rates (condition {\it 4/1})
            for channel 7 and 8. Counting rates are averaged over
            50\,min.

}
\end{figure}

\begin{figure}[H]
\centering
  \mbox{\epsfig{file=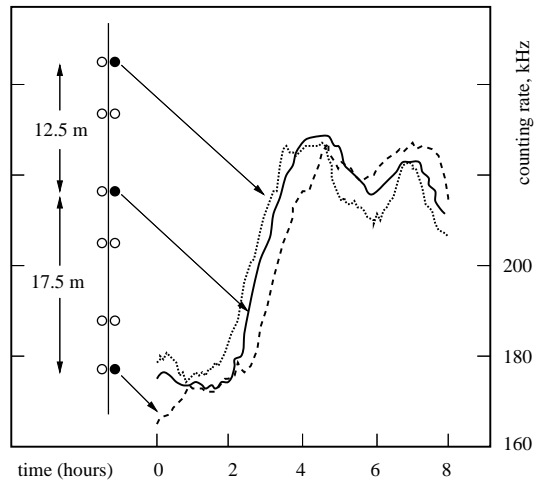,height=7.0cm,angle=-90}}
\caption[2]{\small
Counting rate of three OMs along one string during an
8 hour interval at Sept.\,24, 1993.
}
\end{figure}

\section{Conclusions and Outlook}

The Baikal detector is well understood, and first 
atmospheric neutrinos have
been identified. Their total number agrees well 
with Monte Carlo expectations. Furthermore, 
limits on the flux of GUT magnetic
monopoles have been derived. The detector has the potential
to study so different questions like the search for neutrinos from WIMP
annihilation in the center of the Earth on the one hand,
and monitoring of deep water ventilation 
processes on the other hand. 

The Baikal site is competitive to Oceans: The stronger absorption 
may be taken into account by a somewhat denser spacing 
which, on the other hand,
might  be a natural approach if one focuses to lower thresholds
than in  Oceans. The external
noise is of similar magnitude like in Oceans, but with strong
seasonal variations. The smaller depth has been shown to be
no serious drawback, since with appropriate methods neutrinos
can be separated effectively. The most remarkable advantage of 
the site is the ice cover which allows reliable and inexpensive 
deployment and  retrieval - still an
practically unsolved problem in the case of Ocean projects.

After 144 Optical Modules have been deployed in
March/April 1997,
the {\it NT-200} detector will be completed in April 1998.

In the following years, {\it NT-200} will be operated as a
neutrino telescope with an effective area between 
1000 and 5000 m$^2$ typically, depending on the energy.
This corresponds, after all cuts, to about 1 atmospheric
neutrino per day.
Presumably still too small to detect neutrinos from AGN
and other extraterrestrial sources, {\it NT-200} can be used to push
the flux limits for neutrinos from WIMP annihilation and for
magnetic monopoles. It will also be a unique
environmental laboratory to study water processes
in Lake Baikal.

Apart from its own value, {\it NT-200} is regarded to be a 
prototype for a telescope 20-50 times larger. With
2000 OMs, a threshold of 10-20 GeV and an effective area of
50,000 to 100,000 m$^2$, this telescope would have a
realistic detection potential for extraterrestrial sources
of high energy neutrinos. With its comparatively low threshold,
it would fill a gap between underground detectors and  
planned high threshold detectors of cube kilometer size.

\section{Acknowledgments}
\vspace{-2mm}
This work was supported by the Russian Ministry of Research,the German 
Ministry of Education and Research and the Russian Fund of Fundamental 
Research ( grants {\sf 96-02-17308} and {\sf 97-02-31010}).

\end{document}